\documentclass[twocolumn]{aastex6}

\usepackage{amsmath}
\usepackage{amssymb}
\usepackage{amsthm}
\usepackage{mathrsfs}
\usepackage{bm}
\usepackage{graphicx}
\usepackage{xcolor}
\usepackage[scaled=0.90]{beramono}
\usepackage[T1]{fontenc}
\usepackage[utf8]{inputenc}

\usepackage{hyperref}
\hypersetup{
	colorlinks=true,
	linkcolor={red!50!black},
   citecolor={blue!50!black},
   urlcolor={blue!80!black}	
}

\numberwithin{equation}{section}
\graphicspath{{./Figures/}}

\newcommand{\ve}{\varepsilon}

\newcommand{\Fig}[1]{Fig.~\ref{#1}}

\newcommand{\Sec}[1]{Sec.~\ref{#1}}

\newcommand{\Ms}{M_{\sun}}
\newcommand{\PSRwithMOI}{PSR J0737-3039A}
\newcommand{\constraintPulsar}{B1516+02B}
\newcommand{\EoS}{EoS}
\newcommand{\EoSs}{EoSs}
\newcommand{\fastPulsar}{PSR J1748-2446ad}
\newcommand{\RNS}{RNS}
\newcommand{\cEFT}{ChEFT}
\newcommand{\MoI}{moment of inertia}

\begin{document}

\title{Global properties of rotating neutron stars with QCD equations of state}

\author{Tyler Gorda}
\affil{University of Helsinki, Helsinki, Finland \\ University of Colorado Boulder, Boulder, Colorado}

\date{November 14, 2016}

\begin{abstract}
We numerically investigate global properties of rotating neutron stars using the allowed band of QCD equations of state derived by \citet{Kurkela:2014vha}.  This band is constrained by chiral effective theory at low densities and perturbative QCD at high densities, and is thus, in essence, a controlled constraint from first-principles physics.  Previously, this band of equations of state was used to investigate non-rotating neutron stars only; in this work, we extend these results to any rotation frequency below the mass-shedding limit.  We investigate mass--radius curves, allowed mass--frequency regions, radius--frequency curves for a typical 1.4$\Ms$ star, and the values of the moment of inertia of the double pulsar \PSRwithMOI, a pulsar whose moment of inertia may be constrained observationally in a few years.  We present limits on observational data coming from these constraints, and identify values of observationally-relevant parameters that would further constrain the allowed region for the QCD equation of state.  We also discuss how much this region would be constrained by a measurement of the moment of inertial of the double pulsar \PSRwithMOI.
\end{abstract}

\maketitle


\section{Introduction}

Neutron stars (NSs) are one of the most extreme physical systems in the cosmos.  Within a sphere of radius $\sim$10 km lies over 1$\Ms$ of matter.  In the outer layers of NSs, controlled techniques such as chiral effective theory (\cEFT) \citep{Tews:2012fj} or quantum Monte Carlo \citep{Abbar:2015wda} are applicable and can yield insights into both the static properties of the bulk matter (such as the equation of state or \EoS) and some transport properties. Currently, these low-density calculations are valid up to about 1.1 times the nuclear saturation density $n_{s} \approx 0.16/\text{fm}^{3}$, corresponding to a baryon chemical potential of about $\mu_{B} \approx 0.97$~GeV \citep{Tews:2012fj}.   Deep in the core, however, such controlled, direct theoretical calculations are not possible.  This is because the densities and chemical potentials at the center of the star, though extreme, are not large enough to fall into the range accessible by perturbative quantum chromodynamics (pQCD).  In the state-of-the-art pQCD calculations at zero temperature in \citet{Kurkela:2009gj}, the errors associated with varying the mass scale reach 30\% at around $\mu_{B} = 2.6$\ GeV.  The value of $\mu_{B}$ in  the cores of NSs lie within a subset of this $0.97 - 2.6$~GeV range.

The problem of the interiors of NSs is thus currently a nonperturbative one.  However, one can hope to reach the intermediate values of $\mu_{B}$ by matching the low-density \EoS\ from the low-energy effective theories to the pQCD results in a thermodynamically consistent way to investigate the (static) makeup of NSs.  This has been carried out in the work of \citet{Kurkela:2014vha} and \citet{Fraga:2015xha}, who, in addition, incorporated the 2$\Ms$ constraint from \citet{Demorest:2010bx,Antoniadis:2013pzd}.  (See also \citet{Hebeler:2009iv}, in which the authors use only \cEFT\ and the 2$\Ms$ constraint to extend the low-energy \EoS.) In these works, the authors used their matched \EoSs\ to analyze non-rotating NSs only.  It is known \citep{Benhar:2005gi,Cipolletta:2015nga} that slowly-rotating NSs can be approximated as non-rotating for frequencies of rotation less than about $f \approx 200$\ Hz.  Beyond this, however, one must use numerical codes to analyze the structure of the stars.  Such a numerical approach has been recently used by \citet{Cipolletta:2015nga} and \citet{Haensel:2016pjp} in the context of phenomenological \EoSs.  One of the main purposes of the current article is to extend these analyses to include \EoSs\ that are more fully constrained by first-principles physics.
Broadly speaking, the purpose of this work is to investigate the effects of rotation on NSs all the way up to the mass-shedding limit using the constraints on the QCD \EoS\ determined in \citet{Kurkela:2014vha,Fraga:2015xha}.  We are particularly interested in constraining NS properties that are relevant observationally. As such, we investigate the maximum allowed NS masses, and the allowed regions for mass--radius curves, mass--frequency curves, and radius--frequency curves for a typical 1.4$\Ms$ star.  In addition, we investigate the allowed values of the moment of inertia 
of the double pulsar \PSRwithMOI\ \citep{Kramer:2009zza,Morrison:2004df} and study how this is correlated with the radius.  In this way, we hope that this work will provide the literature with strong direct links between astronomical observations and the allowed QCD \EoSs\ coming from current state-of-the-art pQCD and \cEFT\ calculations.

We note at the outset that even though the pQCD result of \citet{Kurkela:2009gj} assumed local charge neutrality, this does not actually imply that the \EoS\ band of \citet{Kurkela:2014vha} makes this assumption. There are two reasons for this. Firstly, as was noted in \citet{Kurkela:2009gj}, switching between local and global charge neutrality for the pQCD \EoS\ typically leads to a smaller variation in the pressure than is already included in the renormalization scale dependence \citep{Glendenning}.  Secondly, since the polytropic matching carried out in \citet{Kurkela:2014vha} does not preclude the formation of a mixed phase at the matching points (see, e.g., \citet{Glendenning:1992vb}), there is in principle no assumption of local charge neutrality made for the band of \EoSs\ that we use in this work.  

The structure of this paper is as follows:  In \Sec{sec:methodology}, we briefly review the \RNS\ code and describe how it was used to construct all of the aforementioned relations between the NS properties listed above.  In \Sec{sec:results}, we present our results and all of our plots.  In our concluding \Sec{sec:conclusions}, we review our main findings, including ones that are most relevant to astrophysical observation.

\section{Methodology}
\label{sec:methodology}

To conduct our analysis, we used the publicly available \RNS\ code. It can takes as input an \EoS\ in the form $P(\ve)$ and two parameters: a central energy density $\ve$ and the ratio of the polar coordinate radius to the equatorial coordinate radius $r$.  Other inputs can be used as well (see below), but internally each NS that is constructed is specified by the parameters $\ve$ and $r$.  From this input, the code can calculate various global properties of the star, including the total (or gravitational) mass $M$, the circumferential equatorial radius $R_{e}$, the frequency of rotation $f$, and the \MoI\ $I$.  

In addition to constructing a single star specified by $\ve$ and $r$, the \RNS\ code can construct sequences of stars as well as accept other stellar properties as input to construct internal sequences and find stars satisfying those inputs. It can also calculate the mass-shedding frequency for a given central energy density $\ve_{0}$, which is the fastest rotation rate possible before the star begins to throw off mass from its equator. This provides an upper bound on the rotation rate for the central energy density $\ve_{0}$.  Rotating stars have both a larger maximum mass and a larger maximum equatorial radius, and so the mass-shedding limit can be used to investigate larger, more massive stars than were possible in the non-rotating limit.

The approach used in this investigation was to take the \EoSs\ used in \citet{Kurkela:2014vha,Fraga:2015xha} in the form $P(\ve)$ and feed them into the \RNS\ code to calculate various properties of physical interest.  A comment is in order here.  In \citet{Kurkela:2014vha,Fraga:2015xha}, the authors match the \cEFT\ \EoSs\ to the pQCD band using two or three intermediate polytropic \EoSs.  In addition, they perform the matching both with and without latent heat at the matching points of the polytropic \EoSs. In the end, the authors conclude that adding latent heat is actually \emph{more} restrictive on the matching, and, in addition, they found that a third polytrope only minimally increased the range of allowed \EoSs.  In light of these results, we have also only used the bitropic \EoSs\ without latent heat in this work.
\begin{figure*}[t]
\begin{minipage}[t]{\columnwidth}
\includegraphics[width=\textwidth]{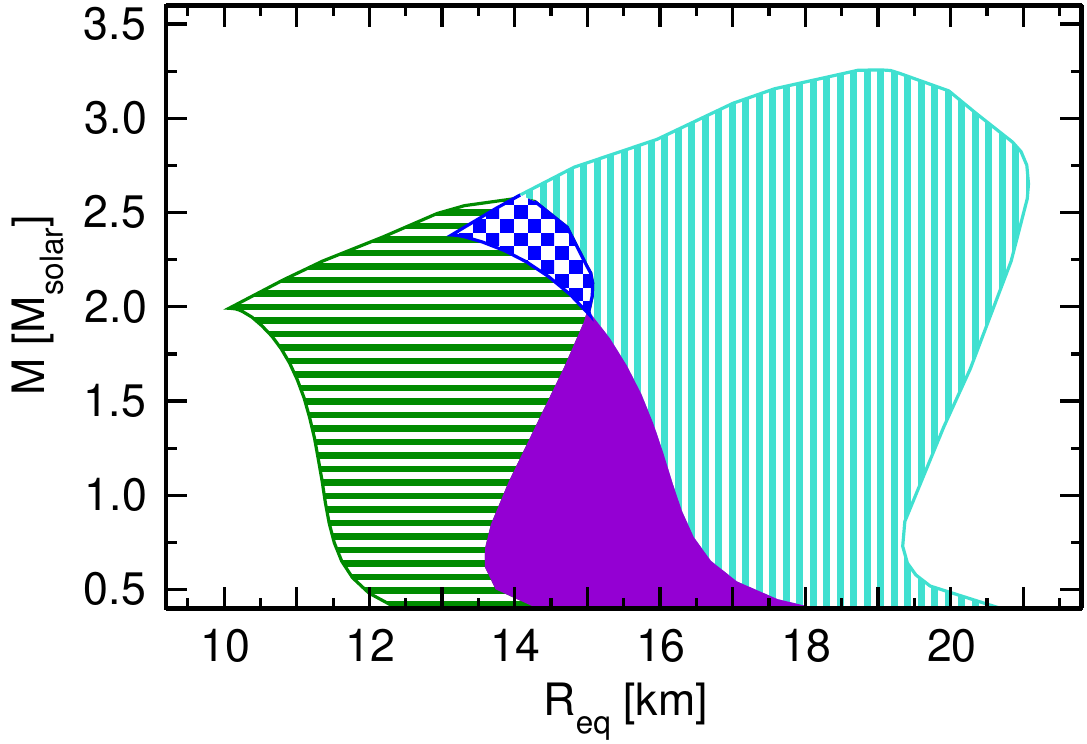}
\caption{(Color online) Mass vs. equatorial radius regions for non-rotating stars (horizontal stripes) and mass-shedding stars (vertical stripes).  The upper, checkered region is an overlap between the non-rotating and mass-shedding regions.  The lower, solid region is only accessible to non-mass-shedding, rotating NSs.}
\label{fig:massRadius}
\end{minipage}
\hfill
\begin{minipage}[t]{\columnwidth}
\includegraphics[width=\textwidth]{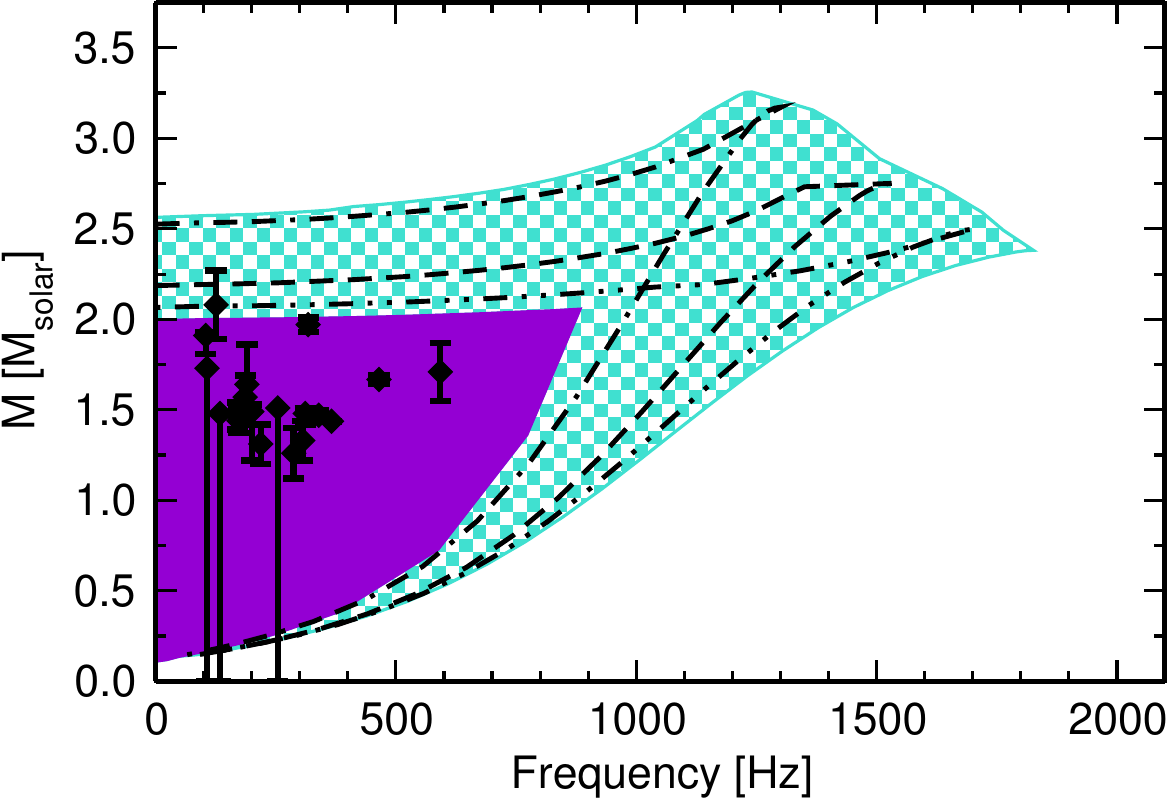}
\caption{(Color online) The allowed mass--frequency region for all of the possible \EoSs.  The inner, solid region is allowed for every equation of state, and the outer, checkered band shows where the possible boundaries are for each \EoS. The dashed lines are the outer boundaries of the mass--frequency regions for three sample \EoSs.  Data points for NSs with $f > 100$Hz, taken from a table in \citet{Haensel:2016pjp}, are also plotted.}
\label{fig:massFrequency}
\end{minipage}
\end{figure*}

To construct our data, we first ran the \RNS\ code on the static and mass-shedding sequences.  From this, we could construct the mass--radius curves and one boundary of the allowed mass--frequency region for NSs.  The rest of our numerical data involved either fixed-frequency runs, fixed-mass runs (or both), or coding a binary search to fill in the gaps where the code was unable to generate the star.  This was necessary in the cases of very small frequencies, as internally the code always uses $r$ as a parameter instead of $f$. (This behavior was also noted in \citet{Cipolletta:2015nga}.)  The fixed-frequency runs were used to determine the other boundary of the allowed mass--frequency region, and the fixed-mass runs were used to determine the radius--frequency relations for a typical, 1.4$\Ms$ NS.  Finally, the fixed mass-and-frequency runs were used for investigating \PSRwithMOI.

\section{Results}
\label{sec:results}

We present first our results for mass vs. equatorial radius curves in \Fig{fig:massRadius}.  The non-rotating region is the same as in \citet{Kurkela:2014vha}, and has a maximum mass of about 2.5$\Ms$.  As seen in the figure, rotating NSs have a larger radius and a larger maximum mass than non-rotating ones.  This can be thought of as a consequence of centrifugal force: the stars with large central energy densities that are unstable past the maximum-mass point for non-rotating stars are stabilized (and their central energy densities are lowered) by the outward centrifugal force in the rotating case.  The larger radius is a consequence of the eccentricity of the star caused by the centrifugal force as well.  We see that the maximum-mass star now has a mass of about 3.25$\Ms$, and the largest star radius is about 21km.

As one might expect, the boundaries of the non-rotating region and the mass-shedding regions in \Fig{fig:massRadius} are formed from the same \EoSs; e.g., the \EoS\ that contains the highest-mass stars in the non-rotating case also contains the highest-mass stars in the mass-shedding case.  This means that any further observational constraints that restrict the left, horizontally-striped region in \Fig{fig:massRadius} will also restrict the right, vertically-striped region in the same way.  

There is available observational data on the correlated masses and radii of some NSs, in particular, those located in low-mass x-ray binaries (for a recent review, see \citet{Ozel:2016oaf}).  Though the uncertainties on these data are sizeable in both mass and radius, considering all of the data together reveals a general region in the mass--radius plane. Moreover, combined analyses can produce still more refined insights.  Comparing our allowed mass--radius regions with the results of the analysis of \citet{Ozel:2015fia}, we see that our non-rotating region (again, the same as that of \citet{Kurkela:2014vha}) fills the larger-radius half of the confidence bands for both the quiescent and thermonuclear-burst data.  Moreover, our non-rotating region intersects with the 68\% confidence bands of every neutron star listed in that work, save one, the quiescent NS labelled ``M28'', and the confidence band for that star only narrowly misses our region on the side of smaller radii. Finally, the astrophysically-inferred mass--radius region presented towards the end of \citet{Ozel:2016oaf} primarily covers a region of slightly smaller radii than our non-rotating results: The authors' ``Astro+Exp'' region just touches the smaller-radius edge of our non-rotating results.  These observations together seem to favor the EoSs of \citet{Kurkela:2014vha} that produce stars with smaller radii.

\begin{figure*}[t]
\begin{minipage}[t]{\columnwidth}
\includegraphics[width=\textwidth]{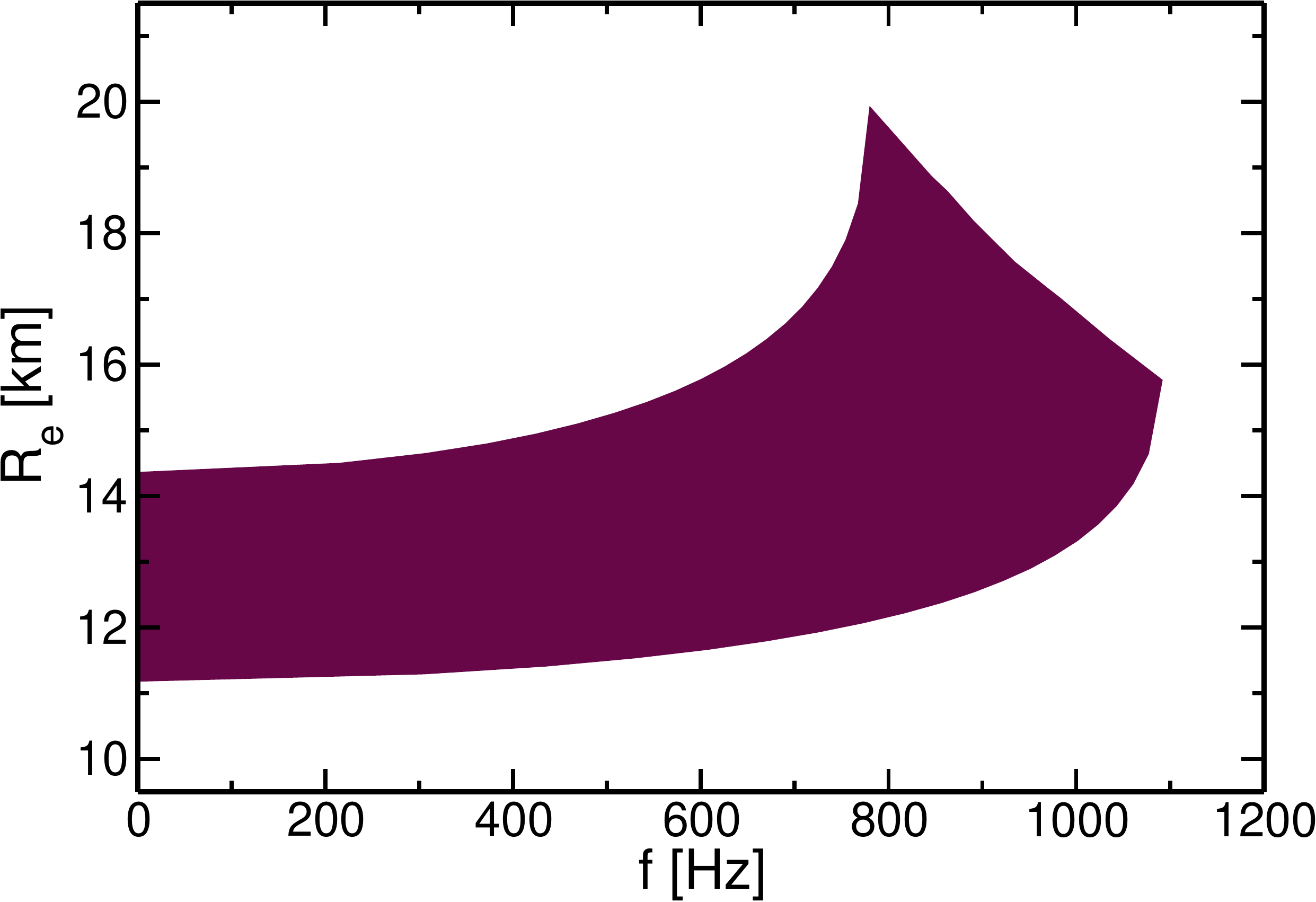}
\caption{The region of allowed circumferential, equatorial radius vs. frequency curves for a 1.4$\Ms$ star.}
\label{fig:radiusFrequency}
\end{minipage}
\hfill
\begin{minipage}[t]{\columnwidth}
\includegraphics[width=\textwidth]{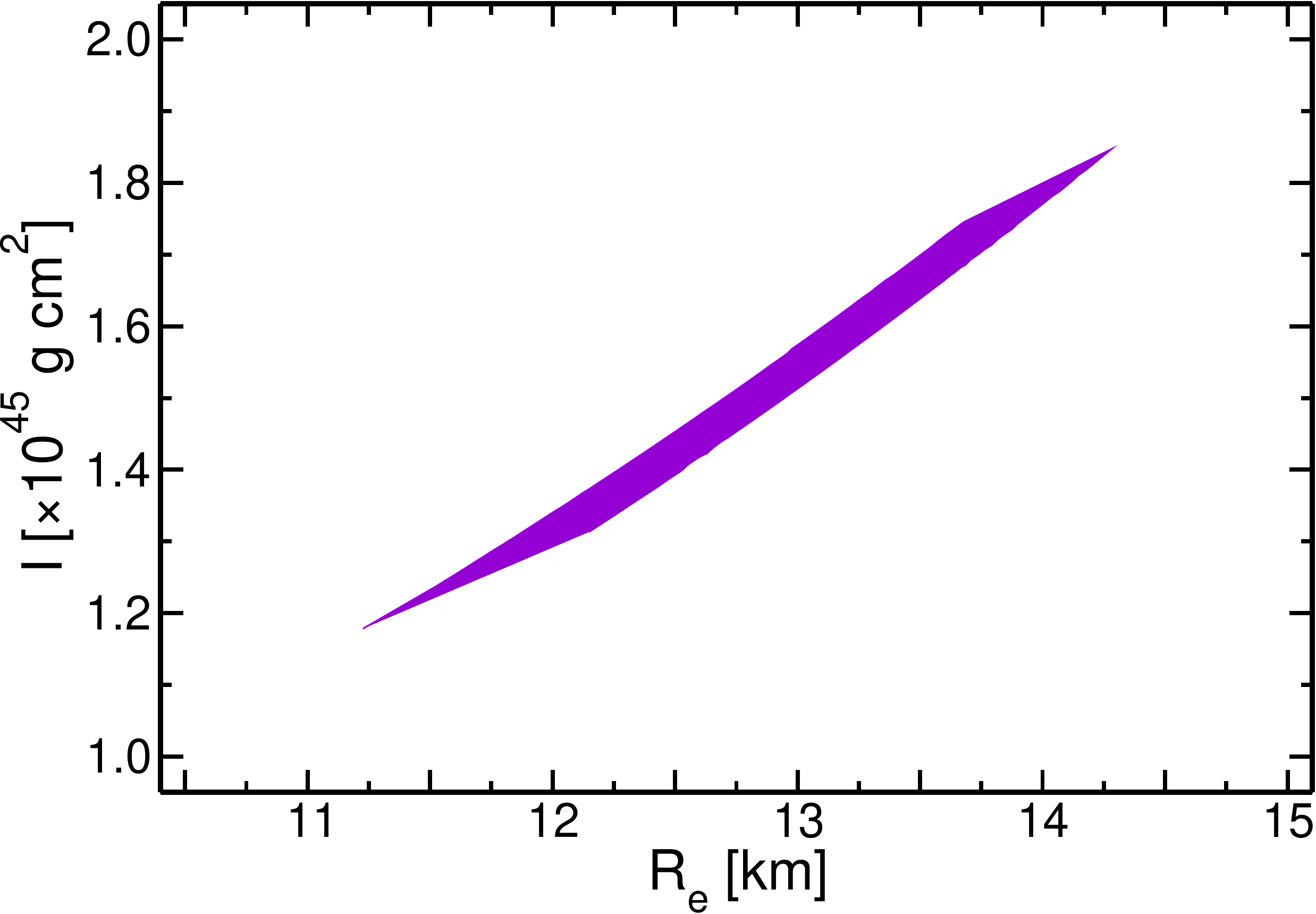}
\caption{The allowed region of \MoI\ vs. circumferential, equatorial radius for \PSRwithMOI.}
\label{fig:radiusMOI}
\end{minipage}
\end{figure*}

In \Fig{fig:massFrequency}, we show the allowed regions for NSs in the mass-frequency plane. The inner, solid region is allowed for every equation of state, and the outer, checkered band shows where the possible boundaries are for each \EoS.  The right boundary of the checkered region is constrained by the mass-shedding stars: beyond a certain limiting frequency at a given mass, stars become unstable.  The upper boundary of the checkered region consists of the curves $M_{\max}(f)$, the maximum NS mass as a function of frequency. We also include three dashed lines in \Fig{fig:massFrequency}, which are the boundaries of the mass--frequency regions for three sample \EoSs.  This is to illustrate the shape of the boundary for each \EoS. Every \EoS\ is shaped similarly to these: the top boundary rises towards the sloped, upper right-hand edge of the checkered region, comes to a point, and then curves back down.  Note that this implies that the outermost boundary of the checkered region is not formed from a single \EoS; in fact, even the upper edge and right edge of the checkered region are formed by different \EoSs.

We also show in \Fig{fig:massFrequency} data points for NSs with frequencies above 100Hz, taken from \citet{Haensel:2016pjp}.  A star located in the checkered band would eliminate some of the \EoSs\ (namely, the ones whose curves in the checkered region are closer to the inner, solid region than the data point of the star).  We see that there is only one star that is pushing into the checkered band: this is \constraintPulsar, with a mass of $2.09 \pm 0.19 \Ms$ \citep{Freire:2007xg}.  If the mass of this star were further constrained, it could potentially eliminate a sizeable number of additional \EoSs.  Note, however, that $f = 125.83$Hz for \constraintPulsar, so this is still within the regime where approximating the star as non-rotating is valid.  Thus, this constraint is not fundamentally one of rotation.  

From \Fig{fig:massFrequency}, however, we see that for high-$f$ stars, there \emph{is} a constraint coming from rotation.  The most clear example of this is the upper-right corner of the inner, solid region with coordinates ($M$, $f$) = ($2.06\Ms$, 883Hz).  This frequency, $f = 883$Hz, signifies the highest frequency that all of the \EoSs\ can support.  Thus, if a NS is ever found with $f > 883$Hz, this would eliminate some of the possible \EoSs\ of \citet{Kurkela:2014vha,Fraga:2015xha}.  We note, however, that this is the highest frequency that would eliminate some \EoSs: lower-frequency NSs could also rule out some \EoSs\ if their masses could be measured and were sufficiently low.  For example, \fastPulsar, currently the fastest rotating NS known ($f = 716$Hz) \citep{Hessels:2006ze}, would eliminate some \EoSs\ if its mass is less than about $1\Ms$. 

For a $1.4\Ms$ NS, the largest frequency that all \EoSs\ can support is lower, $f = 780$Hz, as show in \Fig{fig:radiusFrequency}.  In this figure, we have plotted the equatorial radius as a function of frequency $R_{e}(f)$ for a typical $1.4\Ms$ NS for each \EoS.  This plot serves as a prediction for observational astronomers. Furthermore, when consistent, reliable data of NS radii are available, a plot of this type could be overlaid with observational data to further constrain the QCD \EoS\ (similar to \Fig{fig:massFrequency} above).  One other comment we wish to make here is that this radius--frequency band agrees with the result of the minimum-$\chi^{2}$, hybrid \EoS\ of \citet{Kurkela:2010yk}.  That result lies directly in the center of our band in \Fig{fig:radiusFrequency}.  We do note, however, that their \emph{mass}--frequency boundary only partially agrees with our band: The boundary of the mass--frequency region coming from the mass-shedding curve in \citet{Kurkela:2010yk} lies in the center of our checkered band coming from our mass-shedding curves, but their upper boundary cuts into our solid band.  This is because the minimum-$\chi^{2}$, hybrid \EoS\ obtained in \citet{Kurkela:2010yk} does not permit a 2$\Ms$ NS.

The final plot that we have generated from the \EoSs\ is shown in \Fig{fig:radiusMOI}.  In this figure, we show the allowed region for the \MoI\ and equatorial radius of \PSRwithMOI.  The \MoI\ of this star may be measured in a few years \citep{Kramer:2009zza,Morrison:2004df}, and so it is natural to investigate what the QCD \EoSs\ predicts its value should be.  We find that $I \in [1.2 , 1.8] \times 10^{45}$g~cm${}^{2}$.  Work of this type has been performed previously assuming phenomenological \EoSs, e.g. in \citet{Lattimer:2004nj,Morrison:2004df,Bejger:2005jy}; and, more recently, \citet{Raithel:2016vtt} have performed an analysis in which an \EoS\ is only assumed up to $n_{s}$, and the remaining mass is shifted around to minimize and maximize $I$ for the star.  This allows the authors to plot the largest allowed region in the $R_{e}, I$ plane constrained by controlled, first-principles, low-energy physics.  Our allowed region in \Fig{fig:radiusMOI} does fall within the larger-$R_{e}$, larger-$I$ (i.e., upper-right) portion of the region calculated in the aforementioned work, and it also falls roughly in the center of the forty \EoSs\ data points presented in an earlier figure in that work.  

We also find that all of the ``hard'' and ``soft'' \EoSs\ from \citet{Kurkela:2014vha,Fraga:2015xha} fall on the two boundaries of our allowed region: the ``hard'' \EoSs\ form the right boundary and the ``soft'' ones form the left boundary. In other words, the ``hard'' and ``soft'' \EoSs\ each lie on their own fixed curve.  This is not surprising, since the largest contribution to $I$ comes from the matter at the largest radii (in the low-density crust region), and there, all the ``hard'' or ``soft'' \EoSs\ agree by construction.  But since these \EoSs\ form the vertical boundaries of the region, even a relatively imprecise measurement of the \MoI\ of \PSRwithMOI\ (e.g., one with a precision of 10\%) will \emph{significantly} constrain which \EoSs\ are consistent with the measurement.  Since the allowed region spans $0.6 \times 10^{45}$~g~cm${}^{2}$ in $I$, a 10\% measurement will only be consistent with about $0.15 / 0.6 = 25\%$ of the \EoSs.  

\begin{figure}[t]
\includegraphics[width=0.48\textwidth]{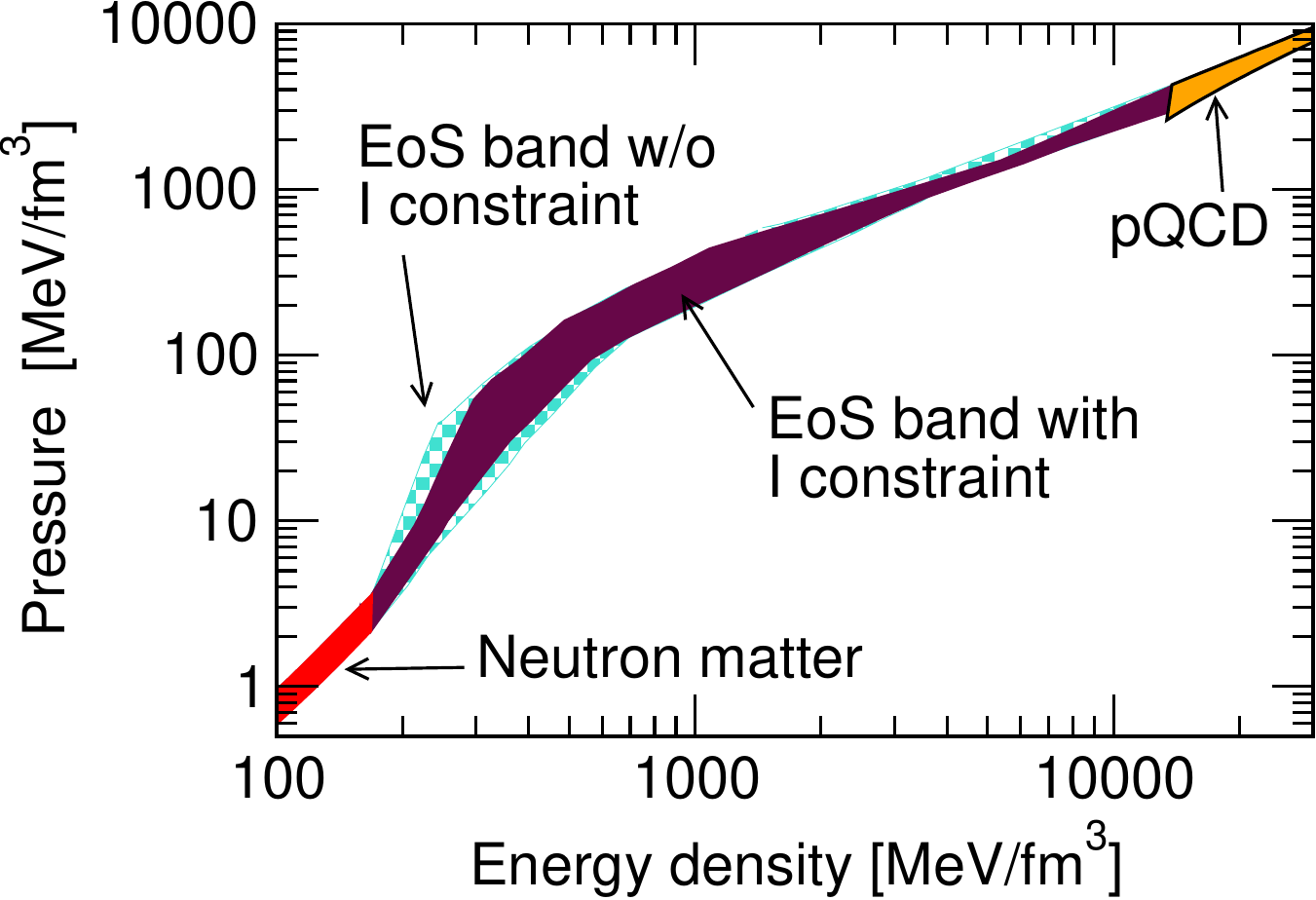}
\caption{(Color online) A plot illustrating how much the \EoS\ band from \citet{Kurkela:2014vha} would be restricted by a hypothetical measurement of $I = 1.5\times 10^{45}$~g~cm${}^{2}$ with a precision of 10\% for \PSRwithMOI.}
\label{fig:reducedBand}
\end{figure}

This percentage is not a physical meaningful result, but we translate it into a statement about the equation of state band in \Fig{fig:reducedBand}.  In this figure, we display the Kurkela et al. \EoS\ band, along with the subset of it that is consistent with $I = 1.5 \times 10^{45}$~g~cm${}^{2}$ to a precision of 10\%, as an example.  We see that such a measurement would shrink the percent errors of the band by up to 50\% in some places, especially in the lowest-density regime.  Again, this makes sense because it is the low-density material farthest from the rotation axis that contributes most to $I$.  This reduction in the \EoS\ band would then, by extension, significantly constrain \emph{all} of the NS properties mentioned in this work.  This makes a measurement of $I$ of the double pulsar \PSRwithMOI\ of extreme interest.  Such a measurement would also constrain the radius of the pulsar to within about $\pm$0.5 km in the context of the \EoSs\ used in this work.

\section{Conclusions}
\label{sec:conclusions}

In this work, we have investigated the effects of rotation on global properties of NSs constructed from the \EoSs\ of \citet{Kurkela:2014vha,Fraga:2015xha}. We have found the maximum allowed NS mass to be about 3.25$\Ms$ and the maximum allowed NS radius to be about 21km.  From investigations of mass--radius relations, we have observed that the smaller-radius results of \citet{Kurkela:2014vha} are favored by current data and analysis.  From investigations of mass--frequency relations, we have have identified \constraintPulsar\ as a NS of particular interest: constraining its mass more precisely could potentially eliminate many allowed QCD \EoSs.  From mass--frequency relations, we also have identified $f = 883$Hz as the maximum allowed NS rotation frequency consistent with every \EoS.  In the case of a canonical 1.4$\Ms$ NS, we have found that $f = 780$Hz is the maximum allowed rotation frequency consistent with every EoS.  We have also determined the allowed $R_{e}$ vs. $f$ region for a 1.4$\Ms$ NS, which may serve has a prediction for astronomers, and may also be overlaid with future precise radius measurements to further constrain the QCD \EoS.  Finally, we have calculated the \MoI\ and radius of \PSRwithMOI\ for each \EoS\ and found it to be consistent with the minimally constrained results of \citet{Raithel:2016vtt}. We have found that $I \in [1.2 , 1.8] \times 10^{45}$g~cm${}^{2}$ for the allowed QCD \EoSs. Most excitingly, we have concluded that even a measurement of the \MoI\ of this star with a precision of 10\% would reduce the percent errors of the band of allowed QCD \EoSs\ consistent with observations to 50\% of its current size at low densities.  We thus conclude that a measurement of the \MoI\ of \PSRwithMOI\ would be of extreme interest.

\acknowledgments
The author would like to thank Aleksi Kurkela, Paul Romatschke, and Aleksi Vuorinen for many helpful discussions and suggestions. 


\bibliography{References}

\begin{thebibliography}{}
\expandafter\ifx\csname natexlab\endcsname\relax\def\natexlab#1{#1}\fi

\bibitem[{Abbar {et~al.}(2015)Abbar, Carlson, Duan, \& Reddy}]{Abbar:2015wda}
Abbar, S., Carlson, J., Duan, H., \& Reddy, S. 2015, Phys. Rev., C92, 045809

\bibitem[{Antoniadis {et~al.}(2013)}]{Antoniadis:2013pzd}
Antoniadis, J., {et~al.} 2013, Science, 340, 6131

\bibitem[{Bejger {et~al.}(2005)Bejger, Bulik, \& Haensel}]{Bejger:2005jy}
Bejger, M., Bulik, T., \& Haensel, P. 2005, Mon. Not. Roy. Astron. Soc., 364,
  635

\bibitem[{Benhar {et~al.}(2005)Benhar, Ferrari, Gualtieri, \&
  Marassi}]{Benhar:2005gi}
Benhar, O., Ferrari, V., Gualtieri, L., \& Marassi, S. 2005, Phys. Rev., D72,
  044028

\bibitem[{Cipolletta {et~al.}(2015)Cipolletta, Cherubini, Filippi, Rueda, \&
  Ruffini}]{Cipolletta:2015nga}
Cipolletta, F., Cherubini, C., Filippi, S., Rueda, J.~A., \& Ruffini, R. 2015,
  Phys. Rev., D92, 023007

\bibitem[{Demorest {et~al.}(2010)Demorest, Pennucci, Ransom, Roberts, \&
  Hessels}]{Demorest:2010bx}
Demorest, P., Pennucci, T., Ransom, S., Roberts, M., \& Hessels, J. 2010,
  Nature, 467, 1081

\bibitem[{Fraga {et~al.}(2016)Fraga, Kurkela, \& Vuorinen}]{Fraga:2015xha}
Fraga, E.~S., Kurkela, A., \& Vuorinen, A. 2016, Eur. Phys. J., A52, 49

\bibitem[{Freire {et~al.}(2008)Freire, Wolszczan, Berg, \&
  Hessels}]{Freire:2007xg}
Freire, P. C.~C., Wolszczan, A., Berg, M. v.~d., \& Hessels, J. W.~T. 2008,
  Astrophys. J., 679, 1433

\bibitem[{Glendenning(2000)}]{Glendenning}
Glendenning, N. 2000, {Compact stars: Nuclear physics, particle physics, and
  general relativity}, 2nd edn. (New York, USA: Springer)

\bibitem[{Glendenning(1992)}]{Glendenning:1992vb}
Glendenning, N.~K. 1992, Phys. Rev., D46, 1274

\bibitem[{Haensel {et~al.}(2016)Haensel, Bejger, Fortin, \&
  Zdunik}]{Haensel:2016pjp}
Haensel, P., Bejger, M., Fortin, M., \& Zdunik, L. 2016, Eur. Phys. J., A52, 59

\bibitem[{Hebeler \& Schwenk(2010)}]{Hebeler:2009iv}
Hebeler, K., \& Schwenk, A. 2010, Phys. Rev., C82, 014314

\bibitem[{Hessels {et~al.}(2006)Hessels, Ransom, Stairs, Freire, Kaspi, \&
  Camilo}]{Hessels:2006ze}
Hessels, J. W.~T., Ransom, S.~M., Stairs, I.~H., {et~al.} 2006, Science, 311,
  1901

\bibitem[{Kramer \& Wex(2009)}]{Kramer:2009zza}
Kramer, M., \& Wex, N. 2009, Class. Quant. Grav., 26, 073001

\bibitem[{Kurkela {et~al.}(2014)Kurkela, Fraga, Schaffner-Bielich, \&
  Vuorinen}]{Kurkela:2014vha}
Kurkela, A., Fraga, E.~S., Schaffner-Bielich, J., \& Vuorinen, A. 2014,
  Astrophys. J., 789, 127

\bibitem[{Kurkela {et~al.}(2010{\natexlab{a}})Kurkela, Romatschke, \&
  Vuorinen}]{Kurkela:2009gj}
Kurkela, A., Romatschke, P., \& Vuorinen, A. 2010{\natexlab{a}}, Phys.Rev.,
  D81, 105021

\bibitem[{Kurkela {et~al.}(2010{\natexlab{b}})Kurkela, Romatschke, Vuorinen, \&
  Wu}]{Kurkela:2010yk}
Kurkela, A., Romatschke, P., Vuorinen, A., \& Wu, B. 2010{\natexlab{b}},
  arXiv:1006.4062, \href{https://arxiv.org/abs/1006.4062}{arXiv:1006.4062}

\bibitem[{Lattimer \& Schutz(2005)}]{Lattimer:2004nj}
Lattimer, J.~M., \& Schutz, B.~F. 2005, Astrophys. J., 629, 979

\bibitem[{Morrison {et~al.}(2004)Morrison, Baumgarte, Shapiro, \&
  Pandharipande}]{Morrison:2004df}
Morrison, I.~A., Baumgarte, T.~W., Shapiro, S.~L., \& Pandharipande, V.~R.
  2004, Astrophys. J., 617, L135

\bibitem[{Ozel \& Freire(2016)}]{Ozel:2016oaf}
Ozel, F., \& Freire, P. 2016, arXiv:1603.02698

\bibitem[{Ozel {et~al.}(2016)Ozel, Psaltis, Guver, Baym, Heinke, \&
  Guillot}]{Ozel:2015fia}
Ozel, F., Psaltis, D., Guver, T., {et~al.} 2016, Astrophys. J., 820, 28

\bibitem[{Raithel {et~al.}(2016)Raithel, Ozel, \& Psaltis}]{Raithel:2016vtt}
Raithel, C.~A., Ozel, F., \& Psaltis, D. 2016, Phys. Rev., C93, 032801,
  [Addendum: Phys. Rev.C93,no.4,049905(2016)]

\bibitem[{Tews {et~al.}(2013)Tews, Krüger, Hebeler, \& Schwenk}]{Tews:2012fj}
Tews, I., Krüger, T., Hebeler, K., \& Schwenk, A. 2013, Phys. Rev. Lett., 110,
  032504

\end{thebibliography}

\end{document}